# Measurement of Inclusive Antiprotons from Au + Au Collisions at $\sqrt{s_{NN}} = 130$ GeV

C. Adler,[11] Z. Ahammed,[23] C. Allgower,[12] J. Amonett,[14] B.D. Anderson,[14] M. Anderson,[5] G.S. Averichev,[9] J. Balewski,[12] O. Barannikova,[9,23] L.S. Barnby,[14] J. Baudot,[13] S. Bekele,[20] V.V. Belaga,[9] R. Bellwied,[31] J. Berger,[11] H. Bichsel,[30] L.C. Bland,[12] C.O. Blyth,[3] B.E. Bonner,[24] A. Boucham,[26] A. Brandin,[18] R.V. Cadman,[1] H. Caines,[20] M. Calderón de la Barca Sánchez,[33] A. Cardenas,[23] J. Carroll,[15] J. Castillo,[26] M. Castro,[31] D. Cebra,[5] S. Chattopadhyay,[31] M.L. Chen,[2] Y. Chen,[6] S.P. Chernenko,[9] M. Cherney,[8] A. Chikanian,[33] B. Choi,[28] W. Christie,[2] J.P. Coffin,[13] T.M. Cormier,[31] J.G. Cramer,[30] H.J. Crawford,[4] M. DeMello,[24] W.S. Deng,[14] A.A. Derevschikov,[22] L. Didenko,[2] J.E. Draper,[5] V.B. Dunin,[9] J.C. Dunlop,[33] V. Eckardt,[16] L.G. Efimov,[9] V. Emelianov,[18] J. Engelage,[4] G. Eppley,[24] B. Erazmus,[26] P. Fachini,[25] V. Faine,[2] K. Filimonov,[15] E. Finch,[33] Y. Fisyak,[2] D. Flierl,[11] K.J. Foley,[2] J. Fu,[15] C.A. Gagliardi,[27] N. Gagunashvili,[9] J. Gans,[33] L. Gaudichet,[26] M. Germain,[13] F. Geurts,[24] V. Ghazikhanian,[6] J. Grabski,[29] O. Grachov,[31] V. Grigoriev,[18] M. Guedon,[13] E. Gushin,[18] T.J. Hallman,[2] D. Hardtke,[15] J.W. Harris,[33] M. Heffner,[5] S. Heppelmann,[21] T. Herston,[23] B. Hippolyte,[13] A. Hirsch,[23] E. Hjort,[15] G.W. Hoffmann,[28] M. Horsley,[33] H.Z. Huang,[6] T.J. Humanic,[20] H. Hümmler,[16] , G. Igo,[6] A. Ishihara,[28] Yu.I. Ivanshin,[10] P. Jacobs,[15] W.W. Jacobs,[12] M. Janik,[29] I. Johnson,[15] P.G. Jones,[3] E. Judd,[4] M. Kaneta,[15] M. Kaplan,[7] D. Keane,[14] A. Kisiel,[29] J. Klay,[15] S.R. Klein,[15] A. Klyachko,[12] A.S. Konstantinov,[22] L. Kotchenda,[18] A.D. Kovalenko,[9] M. Kramer,[19] P. Kravtsov,[18] K. Krueger,[1] C. Kuhn,[13] A.I. Kulikov,[9] G.J. Kunde,[33] C.L. Kunz,[7] R.Kh. Kutuev,[10] A.A. Kuznetsov,[9] L. Lakehal-Ayat,[26] J. Lamas-Valverde,[24] M.A.C. Lamont,[3] J.M. Landgraf,[2] S. Lange,[11] C.P. Lansdell,[28] B. Lasiuk,[33] F. Laue,[2] A. Lebedev,[2] R. Lednický,[9] V.M. Leontiev,[22] M.J. LeVine,[2] Q. Li,[31] S.J. Lindenbaum,[19] M.A. Lisa,[20] F. Liu,[32] L. Liu,[32] Z. Liu,[32] Q.J. Liu,[30] T. Ljubicic,[2] W.J. Llope,[24] G. LoCurto,[16] H. Long,[6] R.S. Longacre,[2] M. Lopez-Noriega,[20] W.A. Love,[2] D. Lynn,[2] R. Majka,[33] S. Margetis,[14] L. Martin,[26] J. Marx,[15] H.S. Matis,[15] Yu.A. Matulenko,[22] T.S. McShane,[8] F. Meissner,[15] Yu. Melnick,[22] A. Meschanin,[22] M. Messer,[2] M.L. Miller,[33] Z. Milosevich,[7] N.G. Minaev,[22] J. Mitchell,[24] V.A. Moiseenko,[10] C.F. Moore,[28] V. Morozov,[15] M.M. de Moura,[31] M.G. Munhoz,[25] G.S. Mutchler,[24] J.M. Nelson,[3] P. Nevski,[2] V.A. Nikitin,[10] L.V. Nogach,[22] B. Norman,[14] S.B. Nurushev,[22] G. Odyniec,[15] A. Ogawa,[21] V. Okorokov,[18] M. Oldenburg,[16] D. Olson,[15] G. Paic,[20] S.U. Pandey,[31] Y. Panebratsev,[9] S.Y. Panitkin,[2] A.I. Pavlinov,[31] T. Pawlak,[29] V. Perevoztchikov,[2] W. Peryt,[29] , V.A. Petrov,[10] E. Platner,[24] J. Pluta,[29] N. Porile,[23] J. Porter,[2] A.M. Poskanzer,[15] E. Potrebenikova,[9] D. Prindle,[30] C. Pruneau,[31] S. Radomski,[29] G. Rai,[15] O. Ravel,[26] R.L. Ray,[28] S.V. Razin,[9,12] D. Reichhold,[8] J.G. Reid,[30] F. Retiere,[15] A. Ridiger,[18] H.G. Ritter,[15] J.B. Roberts,[24] O.V. Rogachevski,[9] J.L. Romero,[5] C. Roy,[26] V. Rykov,[31] I. Sakrejda,[15] J. Sandweiss,[33] A.C. Saulys,[2] I. Savin,[10] J. Schambach,[28] R.P. Scharenberg,[23] N. Schmitz,[16] L.S. Schroeder,[15] A. Schüttauf,[16] K. Schweda,[15] J. Seger,[8] D. Seliverstov,[18] P. Seyboth,[16] E. Shahaliev,[9] K.E. Shestermanov,[22] S.S. Shimanskii,[9] V.S. Shvetcov,[10] G. Skoro,[9] N. Smirnov,[33] R. Snellings,[15] J. Sowinski,[12] H.M. Spinka,[1] B. Srivastava,[23] E.J. Stephenson,[12] R. Stock,[11] A. Stolpovsky,[31] M. Strikhanov,[18] B. Stringfellow,[23] C. Struck,[11] A.A.P. Suaide,[31] E. Sugarbaker,[20] C. Suire,[13] M. Šumbera,[9] T.J.M. Symons,[15] A. Szanto de Toledo,[25] P. Szarwas,[29] J. Takahashi,[25] A.H. Tang,[14] J.H. Thomas,[15] M. Thompson,[3] V. Tikhomirov,[18] T.A. Trainor,[30] S. Trentalange,[6] R.E. Tribble,[27] M. Tokarev,[9] M.B. Tonjes,[17] V. Trofimov,[18] O. Tsai,[6] K. Turner,[2] T. Ullrich,[2] D.G. Underwood,[1] G. Van Buren,[2] A.M. VanderMolen,[17] A. Vanyashin,[15] I.M. Vasilevski,[10] A.N. Vasiliev,[22] S.E. Vigdor,[12] S.A. Voloshin,[31] F. Wang,[23] H. Ward,[28] J.W. Watson,[14] R. Wells,[20] T. Wenaus,[2] G.D. Westfall,[17] C. Whitten, Jr.,[6] H. Wieman,[15] R. Willson,[20] S.W. Wissink,[12] R. Witt,[14] N. Xu,[15] Z. Xu,[2] A.E. Yakutin,[22] E. Yamamoto,[15] J. Yang,[6] P. Yepes,[24] V.I. Yurevich,[9] Y.V. Zanevski,[9] , I. Zborovský,[9] H. Zhang,[33] W.M. Zhang,[14] R. Zoulkarneev,[10] and A.N. Zubarev[9]

(STAR Collaboration)

[1]*Argonne National Laboratory, Argonne, Illinois 60439*
[2]*Brookhaven National Laboratory, Upton, New York 11973*
[3]*University of Birmingham, Birmingham, United Kingdom*
[4]*University of California, Berkeley, California 94720*
[5]*University of California, Davis, California 95616*
[6]*University of California, Los Angeles, California 90095*
[7]*Carnegie Mellon University, Pittsburgh, Pennsylvania 15213*
[8]*Creighton University, Omaha, Nebraska 68178*
[9]*Laboratory for High Energy (JINR), Dubna, Russia*

[10]Particle Physics Laboratory (JINR), Dubna, Russia
[11]University of Frankfurt, Frankfurt, Germany
[12]Indiana University, Bloomington, Indiana 47408
[13]Institut de Recherches Subatomiques, Strasbourg, France
[14]Kent State University, Kent, Ohio 44242
[15]Lawrence Berkeley National Laboratory, Berkeley, California 94720
[16]Max-Planck-Institut fuer Physik, Munich, Germany
[17]Michigan State University, East Lansing, Michigan 48824
[18]Moscow Engineering Physics Institute, Moscow, Russia
[19]City College of New York, New York City, New York 10031
[20]Ohio State University, Columbus, Ohio 43210
[21]Pennsylvania State University, University Park, Pennsylvania 16802
[22]Institute of High Energy Physics, Protvino, Russia
[23]Purdue University, West Lafayette, Indiana 47907
[24]Rice University, Houston, Texas 77251
[25]Universidade de Sao Paulo, Sao Paulo, Brazil
[26]SUBATECH, Nantes, France
[27]Texas A & M, College Station, Texas 77843
[28]University of Texas, Austin, Texas 78712
[29]Warsaw University of Technology, Warsaw, Poland
[30]University of Washington, Seattle, Washington 98195
[31]Wayne State University, Detroit, Michigan 48201
[32]Institute of Particle Physics, Wuhan, Hubei 430079 China
[33]Yale University, New Haven, Connecticut 06520

We report the first measurement of inclusive antiproton production at midrapidity in Au + Au collisions at $\sqrt{s_{NN}} = 130$ GeV by the STAR experiment at RHIC. The antiproton transverse mass distributions in the measured transverse momentum range of $0.25 < p_{\perp} < 0.95$ GeV/$c$ are found to fall less steeply for more central collisions. The extrapolated antiproton rapidity density is found to scale approximately with the negative hadron multiplicity density.

We report the first measurement of inclusive antiproton production at midrapidity in Au + Au collisions at nucleon-nucleon center-of-mass energy of $\sqrt{s_{NN}} = 130$ GeV. The measurement was motivated by the following:

(1) Lattice QCD calculations predict that at sufficiently high energy density matter should be in a state of deconfined quarks and gluons [1]. Large energy densities are expected to give rise to increased production of antibaryons relative to lighter mass particles. For example, a higher temperature in a (locally) equilibrated system would result in a larger relative abundance of antibaryons over pions. At high pion density, multiple-pion fusion into baryon-antibaryon pairs may contribute significantly to the antibaryon yield [2]. Therefore, a measurement of the antiproton yield relative to negatively charged hadrons may provide information about the energy density reached in heavy ion collisions. On the other hand, it has been suggested in the context of the Skyrme model that the baryon-antibaryon production rate can be far above a chemical equilibrium estimate [3]. The inclusive antiproton measurement reported here constitutes an important step toward the goal of understanding the physics of baryon production in heavy ion collisions.

(2) The mechanism of baryon transport over large rapidities has been the focus of theoretical investigations [4,5]. The recent measurement of midrapidity antiproton to proton ($\overline{p}/p$) ratio of 0.6 in central Au + Au collisions at RHIC [6] indicates that a finite net-baryon number is present at midrapidity. This implies that a finite baryon number has been transported over 5 units of rapidity in these collisions. Transport of incoming baryon number over several units of rapidity likely occurs very early in the collision and affects the subsequent evolution of the system [4,5,7]. We extract the net-proton multiplicity density at midrapidity from the inclusive antiproton measurement and the published $\overline{p}/p$ ratio [6].

The measurement reported here was carried out in the summer of 2000 at the Relativistic Heavy Ion Collider (RHIC) by the STAR (Solenoidal Tracker at RHIC) experiment. The STAR detector [8] consists of several detector subsystems in a large solenoid magnet, including a time projection chamber (TPC), a scintillator barrel (CTB), and two zero degree calorimeters (ZDC) [9]. The magnet was operated at 0.25 T. The CTB measured the energy deposited by midrapidity charged particles, and the ZDCs measured beamlike neutrons. The coincidence of the ZDCs formed the experimental minimum bias trigger, and, with an addition of a high CTB signal, provided the central collision trigger.

In the off-line analysis, the collision centrality was determined from the measured charged particle multiplicity in the pseudorapidity range $|\eta| < 0.75$ in the TPC. The multiplicity distribution was subdivided into eight centrality bins [10]. The corresponding fractions of events from the measured minimum bias data sample are tabulated in Table I (first column). The measured minimum bias data sample represented about 80%–90% [10] of the theoretical total hadronic cross section of 7.2 b [11]. For the two most central bins, the central collision trigger events were also included.

Tracks were reconstructed from three-dimensional hits in the TPC. The primary interaction point (primary vertex) was reconstructed from the tracks. Events with a primary vertex within $\pm 30$ cm longitudinally of the TPC center were used in this analysis. Tracks were required to point within 3 cm of the primary vertex (distance of closest approach (DCA) cut) and to have at least 25 (of 45 maximum possible) hits. The antiproton rapidity was limited to $|y| < 0.1$. Reconstructed momentum was corrected for particle energy loss in the detector. Momentum resolution for antiprotons was estimated to be about 2% at a transverse momentum of 0.5 GeV/$c$. The effect of smearing due to the finite momentum resolution on antiproton spectra is negligible.

Particle identification was achieved by the measurement of the truncated mean energy loss, $\langle dE/dx \rangle$, of charged particles in the TPC gas. At a momentum of 0.5 GeV/$c$, the width of the $\langle dE/dx \rangle$ distribution for antiprotons was found to be about 11% for this analysis. We constructed a variable [12], $z = \ln[\langle dE/dx \rangle / \langle dE/dx \rangle_{BB}]$, where $\langle dE/dx \rangle_{BB}$ is a single parameter approximation to the expected Bethe-Bloch value for antiprotons. Figure 1 (left panel) shows $z$ versus transverse momentum $p_\perp$ for negatively charged particles at midrapidity, demonstrating the particle identification capability. We present antiproton results for $p_\perp$ up to 0.95 GeV/$c$. Figure 1 (right panel) shows the $z$ distributions for two $p_\perp$ bins. The $z$ distributions were fitted by the sum of three Gaussians (nine free parameters) corresponding to different particle species. The antiproton raw yield was extracted from the fit results for each centrality and $p_\perp$ bin.

A correction factor was applied to the raw yield to account for losses due to acceptance, tracking inefficiency, and antiproton absorption in the detector. The overall reconstruction efficiency including all these effects was obtained from a full Monte Carlo (MC) simulation, embedding MC tracks into real events on the raw data level and comparing the reconstructed tracks with the input through hit matching in proximity. The overall reconstruction efficiency for the most central collisions is about 80% independent of $p_\perp$ in the range of $0.4 < p_\perp < 0.95$ GeV/$c$ and about 70% at $p_\perp = 0.25$ GeV/$c$. The efficiency increases with decreasing event multiplicity, by about 10% to the most peripheral collisions.

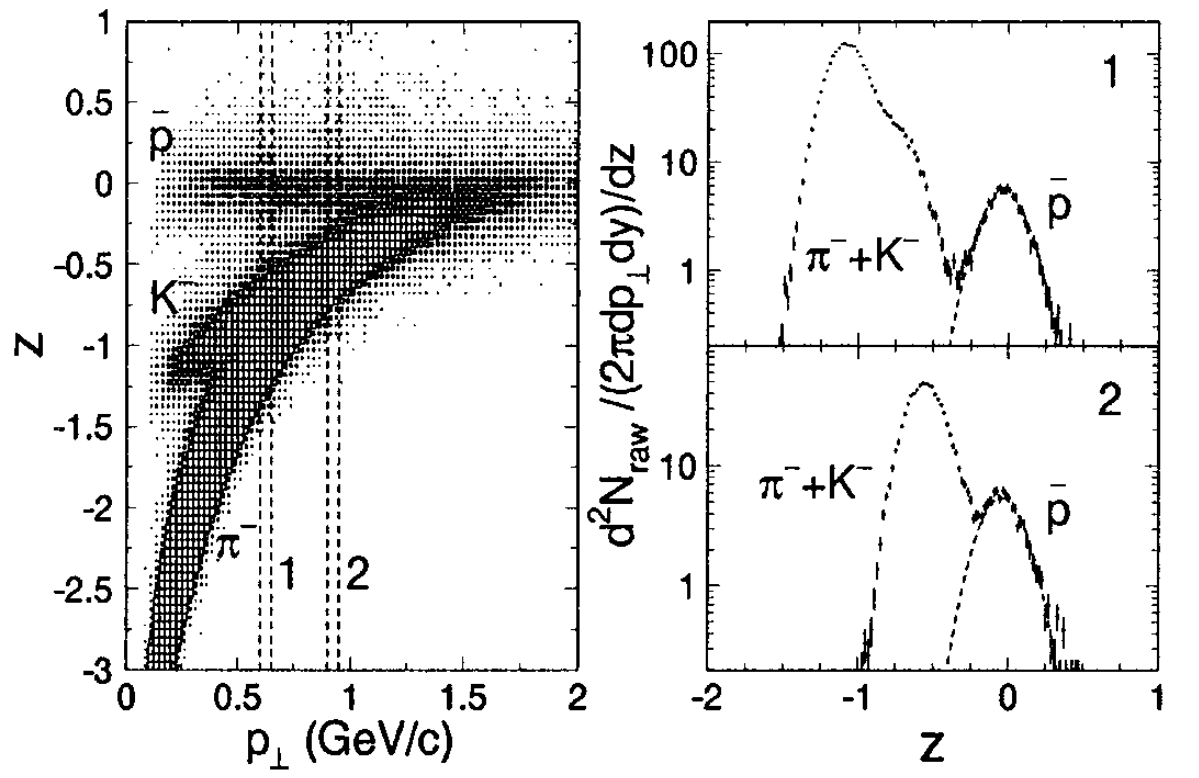

FIG. 1. Left: The $z$ variable versus $p_\perp$ for negatively charged particles. Right: The $z$ distributions for two $p_\perp$ bins, as indicated by the dashed lines in the left panel.

The antiproton yields reported here include secondary products of weak decays. MC studies show that within the DCA cut of 3 cm, the overall reconstruction efficiencies as a function of the measured $p_\perp$ are identical for secondary and primary antiprotons, and the measured antiproton $p_\perp$ spectrum is the sum of the primary antiproton $p_\perp$ spectrum and (0.99 $\pm$ 0.05) times the $p_\perp$ spectrum of secondary antiprotons from weak decays. Secondary antiprotons typically carry most of the parent antihyperon $p_\perp$ due to the decay kinematics, and thus the inclusive antiproton $p_\perp$ distribution is similar to the primary antiproton distribution when the primary antiproton and antihyperon spectra are the same.

Figure 2 shows the antiproton invariant yield per event, $\frac{d^2N}{2\pi m_\perp dm_\perp dy}$, at midrapidity ($|y| < 0.1$) as a function of $m_\perp - m_0$. Here, $m_\perp$ is the transverse mass,

TABLE I. Antiproton fit parameters and yields. Antiprotons are measured at midrapidity ($|y| < 0.1$) and within $0.25 < p_\perp < 0.95$ GeV/$c$. Listed errors are statistical. Systematic errors are 7% on the negative hadron $dN_{h^-}/d\eta$, 10% on the antiproton fiducial $dN/dy$, 15%–25% on the antiproton total $dN/dy$ estimated from the $p_\perp$-Gaussian parametrization, and 10% on $\sigma_{p_\perp}$, $T_{m_\perp}$, and $T_B$.

| Centrality bin | $dN_{h^-}/d\eta$ | Fiducial $dN/dy$ ($0.25 < p_\perp < 0.95$ GeV/$c$) | Total $dN/dy$ ($p_\perp$-Gaussian) | $\sigma_{p_\perp}$ (MeV) | $T_{m_\perp}$ (MeV) | $T_B$ (MeV) |
|---|---|---|---|---|---|---|
| 58%–85% | 17.9 | 0.83 ± 0.02 | 1.15 ± 0.03 | 510 ± 13 | 228 ± 11 | 190 ± 8 |
| 45%–58% | 47.3 | 1.92 ± 0.04 | 2.92 ± 0.09 | 579 ± 17 | 293 ± 17 | 233 ± 11 |
| 34%–45% | 78.9 | 3.00 ± 0.05 | 5.05 ± 0.18 | 642 ± 20 | 362 ± 22 | 274 ± 13 |
| 26%–34% | 115 | 4.04 ± 0.07 | 7.05 ± 0.28 | 668 ± 22 | 390 ± 25 | 290 ± 14 |
| 18%–26% | 154 | 5.29 ± 0.08 | 10.44 ± 0.50 | 741 ± 27 | 481 ± 34 | 338 ± 17 |
| 11%–18% | 196 | 6.41 ± 0.11 | 13.38 ± 0.79 | 777 ± 33 | 525 ± 43 | 359 ± 21 |
| 6%–11% | 236 | 7.84 ± 0.06 | 17.19 ± 0.52 | 804 ± 17 | 568 ± 24 | 378 ± 11 |
| 0%–6% | 290 | 9.49 ± 0.06 | 20.53 ± 0.50 | 799 ± 14 | 560 ± 19 | 374 ± 9 |

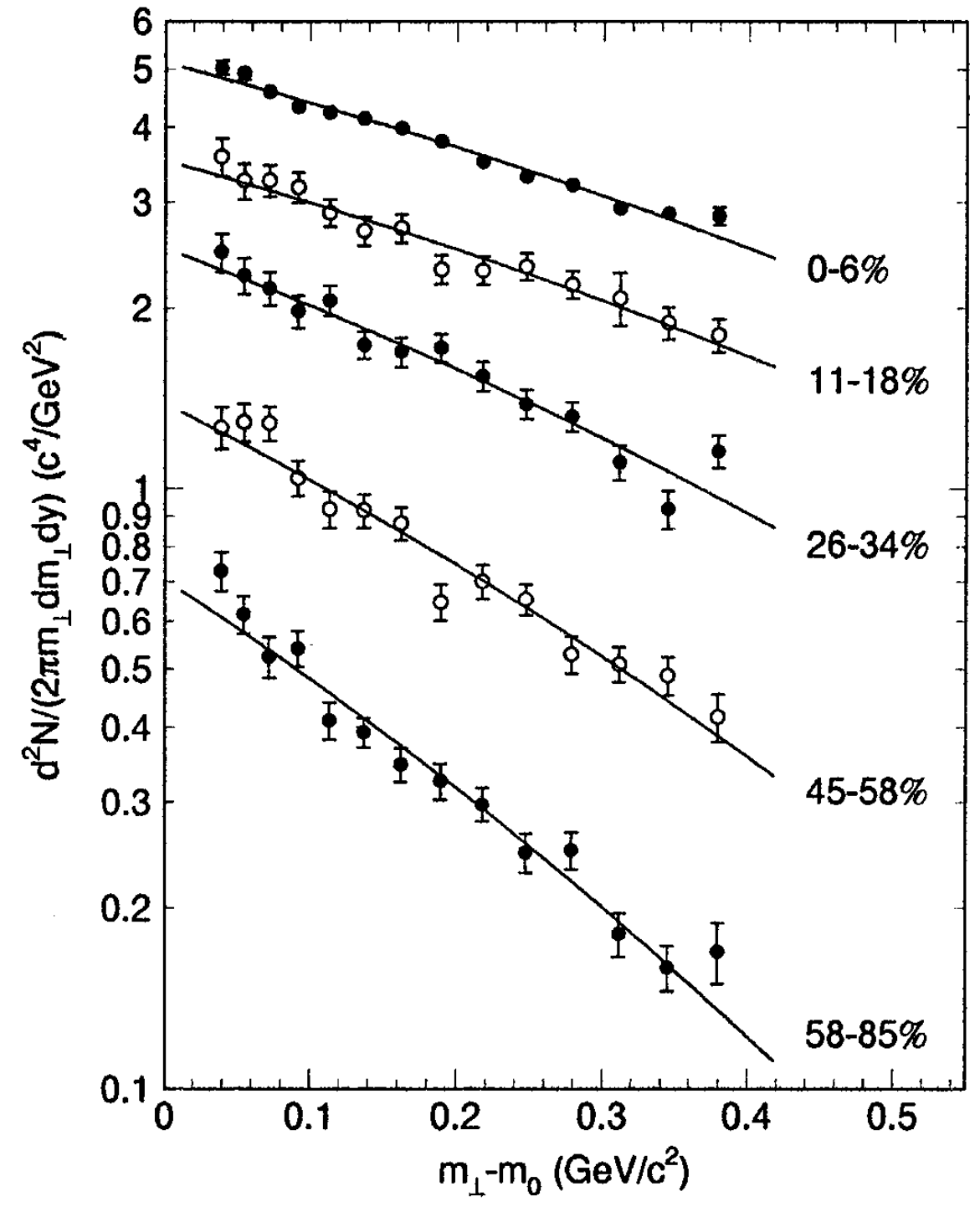

FIG. 2. Transverse mass distributions of inclusive antiproton invariant yield at midrapidity ($|y| < 0.1$). For clarity only five centrality bins are shown. Errors shown are statistical. Systematic errors are 8% point to point and 10% in the overall normalization. The solid lines are $p_\perp$-Gaussian fits to the distributions.

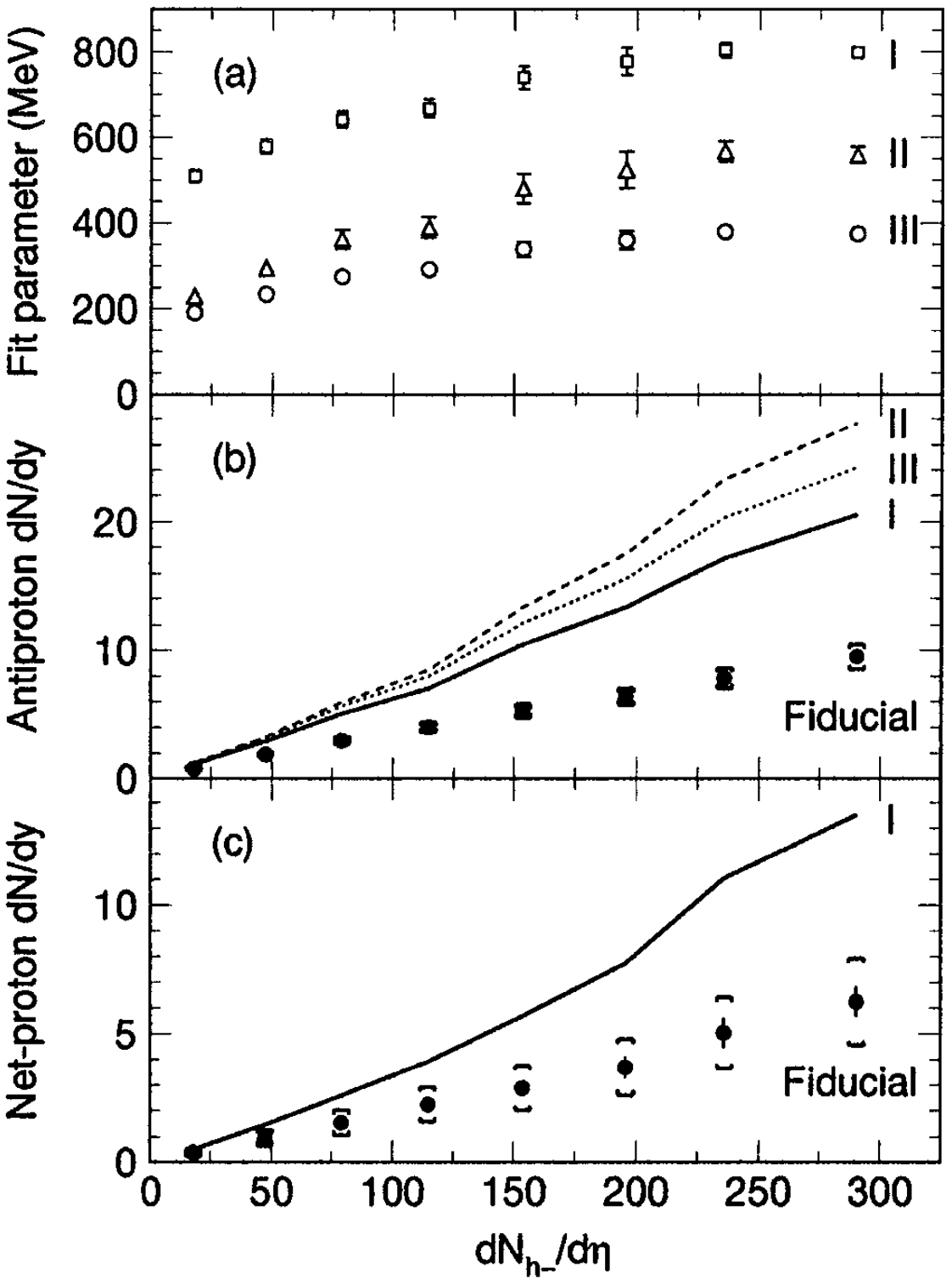

FIG. 3. Dependence of antiproton fit parameters and yields on $dN_{h^-}/d\eta$. (a) Fitted values of parameters from three functional forms described in text: $\sigma_{p_\perp}$ (I), $T_{m_\perp}$ (II), and $T_B$ (III). (b) Fiducial yield of inclusive antiprotons in $0.25 < p_\perp < 0.95$ GeV/$c$ (points) and the estimated total yield using the three functional forms for extrapolation (solid, dashed, and dotted). (c) Fiducial yield of net-protons (points) and the estimated total net-proton yield using $p_\perp$-Gaussian for extrapolation. Error bars are statistical errors. Systematic errors on the fitted parameters in (a) are 10%, and are indicated in "caps" in (b) and (c).

$m_\perp = \sqrt{p_\perp^2 + m_0^2}$, and $m_0$ is the antiproton mass. The uncorrelated point-to-point systematic errors on the spectra were estimated to be 8% by varying the track cuts and by comparing different analysis techniques. Systematic error on the overall normalization was estimated to be less than 10% by varying the event selection as well as the track cuts. As seen in Fig. 2, the spectra in central collisions fall less steeply than in peripheral collisions in the measured $p_\perp$ range of $0.25 < p_\perp < 0.95$ GeV/$c$.

In order to characterize the shape of the spectra quantitatively, we fit the spectra to three functional forms: (I) Gaussian function in $p_\perp$ [$\propto \exp(-p_\perp^2/2\sigma_{p_\perp}^2)$], (II) exponential function in $m_\perp$ [$\propto \exp(-m_\perp/T_{m_\perp})$], and (III) Boltzmann function [$\propto m_\perp \exp(-m_\perp/T_B)$]. Both II and III have been commonly used to characterize $m_\perp$ distributions. The Gaussian function in $p_\perp$ can result from the Schwinger tunneling mechanism for particle production [13,14], but is less commonly used. The three functional forms fit the spectra equally well, with similar $\chi^2$ per degree of freedom of about 1. The $p_\perp$-Gaussian fits are superimposed in Fig. 2. The fit values for $\sigma_{p_\perp}$, $T_{m_\perp}$, and $T_B$, reflecting the local spectra shape, are listed in Table I as a function of the midrapidity negative hadron pseudorapidity density, $dN_{h^-}/d\eta$ [15]. Systematic errors on the fitted values are estimated to be 10% including the effect of the point-to-point systematic errors in the spectra.

Figure 3(a) shows the fitted values for $\sigma_{p_\perp}$, $T_{m_\perp}$, and $T_B$ as a function of $dN_{h^-}/d\eta$ [15], used as a collision centrality estimate. The three parameters clearly exhibit the same monotonic trend with centrality: the more central the collision, the larger the parameters. For comparison, the proton and antiproton parameters, $T_{m_\perp}$, measured in central Pb + Pb collisions at the Super Proton Synchrotron (SPS), are about 300 MeV [16–18]. This value is similar to our result for the peripheral collisions at $dN_{h^-}/d\eta \approx 50$. It has been suggested that the parameters $T_{m_\perp}$ (and similarly $\sigma_{p_\perp}$ and $T_B$), contain information about transverse radial flow which can be generated by a pressure gradient in the collision system [16]. The results indicate qualitatively that at RHIC we observe a stronger transverse flow for midcentral and central events than at SPS.

We characterize the inclusive antiproton production rate by the fiducial rapidity density, $dN/dy$, in the measured $p_\perp$ range, by summing up the data points in each spectrum. The results are listed in Table I and shown in Fig. 3(b) as a function of $dN_{h^-}/d\eta$. In order to estimate the total antiproton rapidity density, we extrapolate our measurement to all $p_\perp$. We show in Fig. 3(b) integrals of the three fitted functional forms. Comparisons with the negative hadron $p_\perp$ distribution [15] reveal that both the $m_\perp$ exponential and the Boltzmann extrapolation of our data for central collisions exceed the negative hadron yield around $p_\perp = 2$ GeV/$c$. Hence, the exponential in $m_\perp$ and the Boltzmann distribution are likely to overestimate the antiproton $dN/dy$. We quote in Table I the $p_\perp$-Gaussian extrapolation as our best estimate of the antiproton total $dN/dy$. By comparing the three integrals, we estimate the systematic errors on the quoted total $dN/dy$ to range from 5% for the most peripheral bin to 15% for the most central bin, in addition to the 10% systematic error on the overall normalization. As seen in Fig. 3(b), the total rapidity density scales approximately with the negative hadron multiplicity. However, as the systematic errors are largely correlated, there is an indication of a larger ratio of the antiproton $dN/dy$ to $dN_{\mathrm{h}^-}/d\eta$ in central collisions than in peripheral collisions.

The centrality dependence of relative antiproton production is qualitatively different from what has been observed at lower energies at the SPS ($\sqrt{s_{NN}} \approx 17$ GeV) [18] and the alternating gradient synchrotron ($\sqrt{s_{NN}} \approx 5$ GeV) [19], where production of antiprotons relative to pions decreases from midcentral to central collisions. Hadronic model studies [20,21] show that this decrease is a result of a strong absorption of antiprotons in the collision zone, but that the initial production of antiprotons relative to pions increases with centrality. We note that absorption of antiprotons may also play a role at the RHIC energy.

A constant $\overline{p}/p$ ratio has been measured at midrapidity in the $p_\perp$ range of $0.4 < p_\perp < 1$ GeV/$c$ [6]. Combining our results with the $\overline{p}/p$ ratio, we extract the midrapidity net-proton ($p - \overline{p}$) fiducial density within $0.25 < p_\perp < 0.95$ GeV/$c$, as shown in Fig. 3(c). The fiducial net-proton density increases approximately linearly with the negative hadron multiplicity. Figure 3(c) indicates that in the most central collisions there are $6.2 \pm 1.8$ protons per unit rapidity in excess of antiprotons in the measured $p_\perp$ range at midrapidity. The total net-proton density at midrapidity can be calculated from the extrapolated antiproton yield and the assumption that $\overline{p}/p$ is constant over all $p_\perp$. This is represented as the curve in Fig. 3(c). The systematic uncertainties on the total net-proton density are 35%, and are largely correlated among centrality bins. Therefore, in central collisions approximately 14 net protons per unit of rapidity are found at midrapidity. For comparison, the HIJING (heavy ion jet interaction generator) model predicts a net-proton density of 6 for central Au + Au collisions at $\sqrt{s_{NN}} = 200$ GeV, while the HIJING model with the baryon junction mechanism predicts a net-proton density of 16 [5,22]. As also seen in Fig. 3(c), the net-proton density exhibits a stronger than linear increase with the negative hadron multiplicity. The nonlinear increase is borne out by the systematic drop of the $\overline{p}/p$ ratio with centrality [6] and the slight increase of the antiproton total rapidity density. The results indicate that more incoming baryons, relative to produced particle multiplicity, are shifted from beam rapidity to midrapidity for more central collisions.

The inclusive antiproton yield reported here is the sum of the primordial antiproton yield and the weak-decay contributions: $\overline{p} + 0.64(\overline{\Lambda} + \overline{\Sigma}^0 + \overline{\Xi} + \overline{\Omega}^+) + 0.52\overline{\Sigma}^-$. We estimate the total antibaryon rapidity density, under the assumption of isospin symmetry with $\overline{n} \approx \overline{p}$ and $\overline{\Sigma}^+ \approx \overline{\Sigma}^0 \approx \overline{\Sigma}^-$, to be twice the measured antiproton rapidity density. Thus, the total net-baryon density is approximately twice the total net-proton density presented in Fig. 3(c).

To conclude, we have measured inclusive antiproton production at midrapidity ($|y| < 0.1$) in the $p_\perp$ range of $0.25 < p_\perp < 0.95$ GeV/$c$ from $\sqrt{s_{NN}} = 130$ GeV Au + Au collisions at RHIC with the STAR experiment. In the measured $p_\perp$ range, the antiproton transverse mass distributions are found to fall less steeply in more central collisions. For the most central collisions, the transverse mass distribution is significantly flatter than in Pb + Pb collisions at the SPS. The antiproton rapidity density at midrapidity is found to scale approximately with the negative hadron multiplicity density. For the most central collisions, over the range $0.25 < p_\perp < 0.95$ GeV/$c$ at midrapidity, we measure $9.5 \pm 1.0$ antiprotons per unit rapidity, resulting in $6.2 \pm 1.8$ protons per unit rapidity in excess of antiprotons taking into account our previous measurement of the $\overline{p}/p$ ratio. From extrapolation of the yields to all $p_\perp$, we find approximately 14 net protons per unit rapidity at midrapidity.

We wish to thank the RHIC Operations Group and the RHIC Computing Facility at Brookhaven National Laboratory, and the National Energy Research Scientific Computing Center at Lawrence Berkeley National Laboratory for their support. This work was supported by the Division of Nuclear Physics and the Division of High Energy Physics of the Office of Science of the U.S. Department of Energy, the United States National Science Foundation, the Bundesministerium fuer Bildung und Forschung of Germany, the Institut National de la Physique Nucleaire et de la Physique des Particules of France, the United Kingdom Engineering and Physical Sciences Research Council, Fundacao de Amparo a Pesquisa do Estado de Sao Paulo, Brazil, and the Russian Ministry of Science and Technology.